\documentclass{ws-procs9x6}

\begin{document}

\title{AN INTRODUCTION TO SOFTWARE ENGINEERING AND FAULT TOLERANCE}

\author{PATRIZIO PELLICCIONE and HENRY MUCCINI}
\address{Dipartimento di Informatica, University of L'Aquila\\
Via Vetoio, 1, 67100 L'Aquila, ITALY\\
E-mail: {pellicci, muccini}@di.univaq.it}

\author{NICOLAS GUELFI}
\address{Laboratory for Advanced Software Systems, University of Luxembourg\\
6, rue Richard Coudenhove-Kalergi, LUXEMBOURG\\
E-mail: nicolas.guelfi@uni.lu}

\author{ALEXANDER ROMANOVSKY}
\address{School of Computing Science, Newcastle University\\
Newcastle upon Tyne, NE1 7RU, UK\\
E-mail: alexander.romanovsky@newcastle.ac.uk}


\begin{abstract}

\end{abstract}

\bodymatter




\section{Motivations for the Book}\label{motivations}

Building systems that are trustful is one of the main challenges
which software developers are facing. Dependability-related concerns
have accompanied system developers since the first day these systems
were built and deployed. Obviously various things have changed since
then, including, the nature of faults and failures, the complexity
of the systems, the services they deliver and the way our society
uses these systems. But the need to deal with various threads (such
as failed components, deteriorating environments, component
mismatches, human mistakes, intrusions and software bugs) is still
in the core of software and system research and development. As
computers are now penetrating various new domains (including the
critical ones) and the complexity of the modern systems is growing,
achieving dependability remains central for system developers and
users.

Accepting that errors always happen in spite of all the efforts to
eliminate the faults that might cause them is in the core of
dependability. To this end various fault tolerance mechanisms have
been investigated by researchers and used in industry.
Unfortunately, more often than not these solutions exclusively focus
on the implementation (e.g. they are provided as middleware/OS
services or libraries) ignoring other development phases, most
importantly the earlier ones. This creates a dangerous gap between
the requirements to build dependable (and fault tolerant) systems
and trying to meet them by exclusively using specific fault
tolerance mechanisms at the implementation step~\cite{crisis}. One
of the consequences of this is that there is a growing number of
reported situations in which fault tolerance means  undermine the
overall system dependability as they are not used properly.

We believe that fault tolerance needs to be explicitly included into
the traditional software engineering theories and practices, and it
should become a part of all steps of software development. As the
current software engineering practices tend to capture only normal
behaviour, assuming that all faults can be removed during
development, new software engineering methods and tools need to be
developed to support explicit handling of abnormal situations.
Moreover, every phase in the software development process needs to
be enriched with the phase-specific fault tolerance means. Generally
speaking, integrating fault tolerance into software engineering
requires:

\begin{itemize}
\item integrating fault tolerance means into system models starting from the
early development phases (i.e. requirement and architecture);

\item making fault tolerance-related decisions for each appropriate model at
each phase by explicit modelling of faults, fault tolerance means and dedicated
redundant resources (with a specific focus on fault tolerant software architectures);

\item ensuring correct transformations of the models used at various development phases
with a specific focus on transformation of the fault tolerance means;

\item supporting verification and validation of the fault tolerance means;

\item developing dedicated tools for fault tolerance modelling;

\item providing domain-specific application-level fault tolerance mechanisms and
abstractions.

\end{itemize}

This book consists of the chapters describing novel approaches to
integrating fault tolerance into software development process. They
cover a wide range of topics focusing on fault tolerance during the
different phases of the software development, software engineering
techniques for verification and validation of fault tolerance means,
and languages for supporting fault tolerance specification and
implementation. Accordingly, the book is structured into the
following three parts:

\begin{itemize}
\item Part A: {\em Fault tolerance engineering: from requirements to code};

\item Part B: {\em Verification and validation of fault tolerant systems};

\item Part C: {\em Languages and Tools for engineering fault tolerant systems}.

\end{itemize}

The next section of this chapter briefly introduces the main
dependability and fault tolerance concepts. Section~\ref{sect1}
defines the software engineering realm. Sections~\ref{fte_sect1},
\ref{VV} and~\ref{languages} introduce the three areas corresponding
to the book parts and briefly outline the current state or research.
The last section summarises the content of the book.
\section{Dependability and Fault Tolerance}\label{dependability}

Dependability is usually defined as the system ability to deliver
service that can be justifiably trusted~\cite{AvizienisLRL04}.
Ensuring the required dependability level for complex computer-based
systems is the challenge which many researchers and developers
working in various relevant domains are facing. The difficulties
here are coming from various sources, including the cost of making
system dependable, the growing complexity of modern applications,
their pervasiveness and openness, proliferation of computer-based
systems into new emerging domains, wider reliance our society puts
on these systems, complexity of ensuring the impact which various
dependability means have on the resulting system dependability,
difficulties in defining realistic and practical assumptions under
which these means are to be applied, difficulties in setting
dependability requirements and tracing them through all development
phases, etc.

Dependability is an integrated concept encompassing a variety of
attributes, including availability, reliability, safety, integrity,
and maintainability. Four general means can be employed to attain
dependability~\cite{AvizienisLRL04}: fault prevention, fault
tolerance, fault removal, and fault forecasting. Clearly in practice
one needs to apply a combination of all of these means to ensure the
required dependability. It is important to understand that all these
activities are centred around the concept of faults where possible
faults are prevented or eliminated by using appropriate development
and verification techniques, the remaining faults are tolerated at
runtime to avoid system failures and estimated to help in predicting
their consequences.

In this chapter, we follow the dependability terminology
from~\cite{AvizienisLRL04} which introduces the following causal
chain of dependability threads. It is said that the system failure
to deliver its service is caused by an erroneous system state,
which, in its turn, is caused by a triggered fault. That means that
faults can be silent for some time and that their triggering does
not necessarily cause immediate failure. Errors are typically latent
and the aim of fault tolerance is to detect them and deal with them
and their causes before they make systems fail.

This book focuses on {\em fault tolerance means that are used to
avoid system failures in the presence of faults}. The essence of
fault tolerance~\cite{Lee} is in detecting errors and carrying the
following system recovery. Generally speaking, during system
recovery one needs to conduct two steps: error handling and fault
handling.

{\em Error handling} can be conducted in one of the following three
ways: backward error recovery (sometimes called rollback), forward
error recovery (sometimes called rollforward) or compensation.
Backward error recovery returns the system into a previous (assumed
to be correct) state. The typical techniques are checkpoints,
recovery points, recovery blocks, conversations, file backup,
application restart, system reboot, transaction abort, etc. Forward
error recovery moves the system into a new correct state, this type
of recovery is typically carried out by employing exception handling
techniques (found, for example, in many programming languages, such
as Ada, Java, C++, etc.). Note that backward error recovery is
usually interpreted as a particular case of forward error recovery.
There have been considerable amount of work on defining exception
handling mechanisms suitable for different domains, development and
modelling paradigms, types of faults, execution environments, etc.
(see, for example, recent book~\cite{Dony}). It is worth noting here
that, generally speaking, the rollforward means are more general and
run-time efficient than the rollback ones as they take advantage of
the precise knowledge of the system erroneous state and move system
into a correct state by using application-specific handlers. To
conduct compensation one needs to ensure that the system contains
enough redundancy to mask errors without interrupting the delivery
of its service.

Various replication and software diversity techniques fall into this
category as they mask the erroneous results without having to move
the system into a state which is assumed to be correct. A wide range
of software diversity mechanisms, including recovery blocks,
conversations and N-version programming, has been developed and
widely used in industry.

{\em Fault handling} activity has a nature which is very different
from error handling as it intends to rid the system from faults to
avoid new errors they may cause in the later execution. It starts
with fault diagnosis, followed by isolation of the faulty component
and system reconfiguration. After that the system or its part needs
to be re-initialized to continue to provide its service. Fault
handling is usually much more expensive than error handling and is
more difficult to apply as it typically requires some part of the
system to be inactive to conduct reconfiguration.

Fault tolerance never comes for free as it always requires
additional (redundant) resources which are employed in runtime for
conducting detection and recovery. Specific fault tolerance
mechanisms require various types of redundancy such as spare time,
additional memory or disk space, extra exchange channels, additional
code or messages, etc. Typically each scheme uses a combination of
redundant resources, for example, simple retry always uses time
redundancy, but may need extra disk space and code to save the
checkpoints if we need to restore the system state before retrying.

The choice of the specific error detection, error handling and fault
handling techniques to be used for a particular system is directly
related to and depends upon the underlying fault assumptions. For
example, replication techniques are typically used to tolerate
hardware faults, whereas software diversity is employed to deal with
software design bugs.

Let us now briefly discuss the main challenges in developing fault
tolerant systems~\cite{sbes}. First of all, the fault tolerance
means are difficult to develop or to use - this is because they
increase system complexity by adding a new dimension to the
reasoning about system behaviour. Their application requires a deep
understanding of the intricate links between normal and abnormal
behaviour and states of systems and components, as well as the state
and behaviour during recovery. Secondly, fault tolerance (e.g.
software diversity, rollback, exception handling) is costly as it
always uses redundancy. Thirdly, system designers are typically
reluctant to think about faults at the early phases of development.
This results in making earlier decisions ignoring fault tolerance,
which may make it more difficult or expensive to introduce fault
tolerance at the later phases. More often than not, the developers
fail to apply even the basic principles of software fault tolerance.
For example, there is no focus on (i) a clear definition of the
fault assumptions as the central step in designing any fault
tolerant system, (ii) developing means for early error detection,
(iii) application of recursive system structuring for error
confinement, (iv) minimising and ensuring error confinement and
error recovery areas, and (v) extending component specification with
a concise or complete definition of failure modes. We can refer here
to recent paper reporting a high number of mistakes made in handing
exceptions in the C programs~\cite{idiom} and to the Interim Report
on Causes of the August 14th 2003 Blackout in the US and
Canada~\cite{blackout}. which clearly shows that the problem was
mostly caused by badly designed fault tolerance: poor diagnostics of
faults, longer-than-estimated time for component recovery, failure
to involve all necessary components in recovery, inconsistent system
state after recovery, failures of alarm systems, etc. It is worth
reminding here, as well, that the failure of the Ariane 5 launcher
was caused by improper handling of an exception~\cite{ariane}.

All the factors above contribute to the fact that a substantial part
of system failures are caused by mistakes in fault tolerance
means~\cite{crisis}. We believe that a closer synergy between
software engineering phases, methods and tools and fault tolerance
will help alleviating such current problems.

\section{Defining Software Engineering}\label{sect1}


Software engineering (SE) is a quite new field of Computer Science,
recognized in the 1968 NATO conference in Garmisch (Germany) as an
emergent discipline. Today, many different definitions of software
engineering have been proposed, trying to explain its main
characteristics:

\begin{itemize}
\item ``Application of {\em systematic, disciplined, quantifiable approach} to
the {\em development, operation, and maintenance} of
software"~\cite{IEEE1471};

\item ``Software engineers should adopt a {\em systematic and organised
approach} to their work and use appropriate tools and techniques
depending on the problem to be solved, the development constraints
and the resources available"~\cite{Sommerville};

\item ``Software Engineering is the field of computer science that
deals with the building of software systems that are {\em so large
or so complex} that they are built by a team or teams of
engineers"~\cite{GhezziJazayeriMandrioli};

\item ``Software engineering is the branch of systems engineering
concerned with the development of {\em large and complex software
intensive systems}". ... ``It is also concerned with the {\em
processes, methods and tools} for the development of software
intensive systems in an {\em economic and timely
manner}"~\cite{EmmerichBook}.
\end{itemize}


Most of those well known definitions point out different
characteristics or perspectives to be used when looking at the
software engineering discipline. Some examples, taken from
experienced computer scientists made in the last forty years, will
help to identify the key points common to most of the definitions of
SE, and help us to illustrate why and when the software engineering
discipline is needed. The $i)$ Ariane 5, the Therac-25 radiation
therapy machine, the Denver Airport (and others) big software
failures~\cite{FatalDefect}, and the $ii)$ on-board shuttle group
excellence~\cite{Fishman96} examples will be used for this purpose.

\subsection{If Software Fails, It May Cost Millions of Dollars and
Injure People}\label{sect11}

As already pointed in Section~\ref{motivations}, software is
pervasive (it is everywhere around us, even if we do not see it), it
controls many devices used everyday, and more and more critical
systems (i.e., those systems whose malfunctioning can injure people
or create high economic losses).

The Ariane 5, X-ray machine and Denver Airport are some examples of
critical systems which, due to software systems malfunctioning,
ended up being big catastrophic failures\footnote{Those and many
other examples of catastrophic failures are described in
~\cite{FatalDefect}.}. The Ariane 5 shuttle, launched on June 4th
1996, broke up and exploded forty seconds after initiation of the
flight sequence, due to a software problem. People were killed. The
Therac-25 radiation-treatment machine for cancer therapy injured and
even killed several patients by administering a radiation overdose.
The Denver Airport software, responsible for controlling 35
kilometers of rails and 4000 tele-wagons never worked properly and
after 10 years of recurring failures it has been recently
dismissed~\cite{JacksonScientificAmerican}: millions of dollars were
wasted. In all three cases, the main causes of failure were
undisciplined management of requirements, imprecise and ambiguous
communication, instable architectures, high complexities,
incoherence among requirements design and implementation, low
automation, and insufficient verification and validation.

\subsection{How to Make Good Software}\label{sect12}

While the previous examples described what might happen when
software engineering techniques are not taken into consideration,
the on-board shuttle group example of excellence (taken from the
1996 white-paper written by Fishman~\cite{Fishman96}) shows results
that can be achieved when software engineering best practices are
applied in practice. It describes how the software governing a
120-ton space shuttle is conceived: such a software system is
composed of around 500.000 lines of code, it controls 4 billion
dollars worth of equipment, and decides the lives of a half-dozen
astronauts. What makes this software and their creators so
extraordinary, is that it never crashes and is bug free (according
to~\cite{Fishman96}). The last three versions of the program (each
one of 420,000 lines of code) had just one fault each. The last 11
versions of this software system had a total of 17 faults.
Commercial programs of equivalent complexity would have 5,000
faults. How did the developers achieve such high software quality?

It was simply the result of applying most of the Software
Engineering (SE) best practices:

\begin{itemize}
\item {\em SE allows for a disciplined, systematic, and quantifiable
development:} The on-board shuttle group is the antithesis of the
up-all-night, pizza-and-roller-hockey software coders who have
captured the public imagination. To be this good, the on-board
shuttle group is very ordinary -- indistinguishable, focused,
disciplined, and methodically managed creative enterprise;

\item {\em SE does not only concern programming:}
Another important factor discussed in the Fishman's
report~\cite{Fishman96} is that about one-third of the process of
writing software happens before anyone writes a line of code. Every
critical requirement is documented. Nothing in the specification is
changed without agreement and understanding. No coder changes a
single line of code without carefully outlining the change;

\item {\em SE takes into consideration maintenance and evolution:}
As explicitly stated in~\cite{IEEE1471} maintenance and evolution
are important factors when engineering software systems. They allow
system evolution, while limiting newly introduced faults;

\item {\em SE is for mid to large systems:}
Applying SE practices is indeed expensive and requires effort. While
non critical, small systems can require just a few SE principles,
applying the best SE practices for the development of critical,
large software systems is mandatory;


\item {\em Development cost and time are key issues:}
The main success of this example is not the software but the
software process the team uses. Recently, much effort has been spent
on identifying new software processes (like the Unified Software
Process~\cite{UmlProcess}), and software maturity frameworks which
allows the improvement of the software development process (like the
Capability Maturity Model -- CMM~\cite{CMM} or the Personal Software
Process -- PSP~\cite{PSP}). Nowadays software processes take
explicitly into consideration tasks like managing groups, setting
deadlines, checking the system cost to stay on budget, and to
deliver software which respects the expected qualities.

\end{itemize}

\section{Fault Tolerance Engineering: from Requirements to
Code}\label{fte_sect1}



Initial solutions relegated fault tolerance (and specifically,
exception handling) very late during the design and implementation
phases of the software life-cycle. More recently, instead, the need
of explicit exception handling mechanisms during the entire life
cycle has been advocated by some researchers as one of the main
approaches to ensure the overall system
dependability~\cite{RomLemos01,ExceptionHandling_SPE05}.

In particular, it has been recognized that different classes of
faults, errors and failures can be identified during different
phases of software development. A number of studies have been
conducted so far aiming to understand where and how fault tolerance
can be integrated in the software life-cycle.

In the remaining part of Section~\ref{fte_sect1} we will show how
fault tolerance has been recently addressed at the different phases
of the software process. The phases that will be taken into
consideration are requirements, high-level (architectural) design,
and low-level design thus reflecting current study on fault
tolerance techniques during such phases.

\subsection{Requirements Engineering and Fault Tolerance}\label{fte_sect11}

Requirements Engineering is concerned with identifying the purpose
of a software system, and the contexts in which it will be used.
Different theories and methodologies for finding out, modelling,
analyzing, evolving and checking software system
requirements~\cite{FOSErequirements} have been proposed so far.

Being requirements the first artefacts produced during the software
process, it is important to document expected faults and ways to
tolerate them. Some approaches have been proposed for this purpose,
the most known being analyzed
in~\cite{Kienzle_Models2005,ExceptionHandling_SPE05,Kienzle_Models2006}
and subsequent work.

In~\cite{Kienzle_Models2005,KienzleSpringer06,KienzleSpringer06} the
authors describe a process for systematically investigating
exceptional situations at the requirements level and provide an
extension to standard UML use case diagrams in order to specify
exceptional behaviour. In~\cite{ExceptionHandling_SPE05} it is
described how exceptional behaviours can be specified at the
requirements level, and how those requirements can drive
component-based specification and design according to the Catalysis
process. In~\cite{Kienzle_Models2006} an approach for analyzing the
safety and reliability of requirements based on use cases is
proposed: normal use cases are extended with exceptional use cases
according to~\cite{Kienzle_Models2005}, then use cases are annotated
with their probability of success and successively translated into
Dependability Assessment Charts, eventually used for dependability
analysis.


\subsection{Software Architecture and Fault Tolerance}\label{fte_sect12}

Software Architecture (SA) has been largely accepted as a well
suited concept to achieve a better software quality while reducing
the time and cost of production. In particular, a software
architecture specification~\cite{SAwolf} represents the first, in
the development life-cycle, complete system description. It provides
both a high-level behavioural abstraction of components and of their
interactions (connectors) and, a description of the static structure
of the system.

Typical SA specifications model only {\em normal} behaviour of the
system, while ignoring {\em abnormal} ones. As a consequence, the
system may fail in unexpected ways due to some faults. In the
context of critical systems with fault tolerance requirements it
becomes necessary to introduce fault tolerance information at the
software architecture level. In fact, the error recovery
effectiveness is dramatically reduced when fault tolerance is
commissioned late in the software life-cycle~\cite{RomLemos01}.

Many approaches have been proposed for modelling and analyzing fault
tolerant software architectures. While a comprehensive survey on
this topic is described in~\cite{MucciniRomanovsky07}, this
introductory section simply identifies the main topics covered by
the existing approaches, while providing some references to the
existing work.

\begin{itemize}
\item Fault Tolerant SA specification: as discussed in many papers
(e.g.,~\cite{ADL_Neno00,NenoUML02,GarlanFSM03}) a software
architecture can be specified using box-and-line notations, formal
architecture description languages (ADLs) or UML-based notations. As
far as the specification of fault tolerant software architectures is
concern, both formal and UML-based notations have been used. The
approaches proposed
in~\cite{CastorFilhoLADC03,WADS05_SAExceptionHandling,GacekDeLemos_BC_CBSIP06}
are examples of formal specifications for Fault Tolerant SA:
traditional architecture description languages are usually extended
in order to explicitly specify error and fault handling. The
approaches in, e.g.,
\cite{DRIP_Catalyst,ExceptionHandling_SPE05,Brito_LADC05} use
UML-based notations for modelling Fault Tolerant SA: new UML
profiles are created in order to be able to specify fault tolerance
concepts;d

\item Fault Tolerant SA analysis: analysis techniques (such as
deadlock detection, testing, checking, simulation, performance)
allow software engineers to assess the software architecture and to
evaluate its quality with respect to expected requirements. Some
approaches have been proposed for analyzing Fault Tolerant SA: most
of them check the architectural model conformance to fault tolerant
requirements or constraints (like
in~\cite{CastorFilho_REFT05,DeLemosWADS2006}). A testing technique
for Fault Tolerant SA is presented in~\cite{Brito_LADC05};

\item Fault Tolerant SA styles: according to~\cite{SC97}, an
architectural style is ``{\em a set of design rules that identify
the kinds of components and connectors that may be used to compose a
system or subsystem, together with local or global constraints on
the way the composition is done}". Many architectural styles have
been proposed for Fault Tolerant SA: the idealized fault tolerant
style (in~\cite{Brito_LADC05}), the iC2C style (which integrates the
C2 architectural style with the idealized fault tolerant component
style~\cite{CastorFilhoLADC03}), the idealized fault tolerant
component/connector style~\cite{DeLemos_IEEESoftw06};

\item Fault Tolerant SA middleware support: when coding software architecture
via component-based systems, middleware technology can be used to
implement connectors, coordination policies and many other features.
In~\cite{Valerie_HICSS01} a CORBA implementation of the proposed
architectural exception handling is proposed.
In~\cite{SEM05_SAExceptionHandling} the authors propose an approach
for exception handling in component composition at the architectural
level with the support of middleware. Many projects have been
conducted to provide fault tolerance to CORBA applications, like
AQuA, Eternal, IRL, and OGS (see~\cite{Bondavalli_IEEETDSC04}). More
details about middlewares are given in Section~\ref{languages}.

\end{itemize}

\subsection{Low-level Design and Fault Tolerance}\label{fte_sect13}

The low-level design phase (hereafter simply called ``design") takes
as input information collected during the requirement and
architecting phases and produces a design artefact to be used by
developers for guiding and documenting software coding. When dealing
with fault tolerant systems, the design phase needs to benefit from
some clear and domain specific tools and methodologies to drive the
implementation of a particular fault tolerant technique. This
introductory section will present only such approaches that treat
fault tolerant during the software development process (from
requirements to code).

In~\cite{Beder_ISORC01,GarciaRubira_BC_AEHT01} two approaches are
presented for transiting from architectural design to low-level
design through the definition of fault tolerant design patterns.
In~\cite{DRIP_Catalyst} an initial study on the CORRECT MDA approach
is introduced: given a coordinated atomic action specification, the
approach enables the automatic production of Java code. This
approach has been successively refined in other papers, and recently
presented in~\cite{Capozucca06}. In~\cite{ExceptionHandling_SPE05}
an approach for fault tolerance specification and analysis during
the entire development process is proposed. It considers how to
specify normal and exceptional requirements, how to use this
information for driving the component specification and design
phase, and how to implement all such information using a Java-based
framework Java. The proposed software process is based on the
Catalysis process. In~\cite{Brito_LADC05} a similar strategy is
adopted based on the UML Components Process (even if this paper is
more towards testing).


\section{Verification and Validation of Fault Tolerant Systems}\label{VV}

Fault tolerance techniques alone are not enough to achieve full
dependability, since unexpected faults cannot be always avoided nor
tolerated~\cite{FT+Surviv99}. In addition it is important to note
that fault tolerant systems inevitably contain faults. Verification
and validation (V\&V) techniques are demonstrated successful means
to assure that expected properties and requirements are satisfied in
system models and implementation~\cite{Brito_LADC05}. In this
setting V\&V techniques are the solution for removing faults from
the system.

Furthermore, V\&V should be used at each different life-cycle phase
since fault tolerance engineers the entire software development
life-cycle.




Different classes of faults, errors, and failures must be identified
and dealt with at each phase of software development, depending on
the abstraction level used in modeling the software system under
development. Thus, each abstraction level requires specific design
models, implementation schemes, verification techniques, and
verification environments.

Verification and validation techniques aim to ensure the correctness
of a software system or at least to reduce the number or severity of
faults both during the development and after the deployment of a
system. There are two different class of verification methods,
exhaustive methods that conduct an exhaustive exploration of all
possible behaviours and non-exhaustive methods that explore only
some of the possible behaviours. In the exhaustive class there are
model checking, theorem provers, term rewriting systems, proof
checker systems, and constraint solvers. In the non-exhaustive class
there are testing and simulation, the veteran techniques, commonly
and widely used but that can easily miss significant errors when
verifying complex and huge systems. In literature several approaches
have been proposed in the last years trying to apply V\&V techniques
to fault-tolerant systems and they are surveyed in the following
subsections. In particular, Section~\ref{sec:MC} reports the use of
model checking techniques for fault-tolerant systems,
Section~\ref{sec:TP} summarizes experiences with theorem provers
applied to fault tolerant systems, Section~\ref{sec:CS} shows
approaches that exploit constraint solvers for verifying fault
tolerant systems, and finally Section~\ref{sec:testing} reports how
fault tolerance has been complemented with testing techniques.
Furthermore, with the introduction of UML~\cite{UML} as the de-facto
standard to model software systems and its widespread adoption in
industrial contexts, many approaches have been proposed to use UML
for modeling and evaluating dependable systems (e.g.,
\cite{ganesh2002,bondavalli99,BondavalliCLMPS01,MPB03}); they are
reported in Section~\ref{sec:UML}.

\subsection{Model Checking}\label{sec:MC}

Model checkers take as input a formal model of the system, typically
described by means of state machines or transition systems, and
verify if it satisfies temporal logic properties~\cite{MCbook}.

Several approaches have been proposed in the last years focusing on
model checking of fault tolerant systems, such
as~\cite{fantechi,autVerFaulTolerance,MCFT}.

These papers describe approaches and show their application to real
case studies testifying that model checking is a promising and
successful verification technique. First of all model checking
techniques are supported by tools, which facilitates their
application. Secondly, in case the verification detects a violation
of a desired property, a counter example showing how the system
reaches the erroneous state in which the property is violated is
produced.

Model checking approaches for fault tolerant systems typically
require to specify normal behaviours, failing behaviours, and fault
recovering procedures. Thus, fault tolerant systems are subjected to
the state explosion problem that afflicts model checkers also in
verifying systems that do not consider exceptional behaviours.

One approach that can be used for avoiding the state explosion
problem is the partial model checking technique introduced
in~\cite{PMC}. This technique that tries to gradually remove parts
of the system is successfully applied for security analysis and an
attempt to use it for fault tolerant systems is in~\cite{GLM03:svv}.

\subsection{Theorem Provers}\label{sec:TP}

Interactive theorem provers start with axioms and try to produce new
inference steps using rules of inference. They require a human user
to give hints to the system. Working on hard problems usually
requires a skilled user. A logic characterization of fault tolerance
is given in~\cite{GLM03:svv}, while approaches that apply theorem
prover techniques to fault tolerant systems are
in~\cite{TePro,abstraction,PVS2006,PVSFTArch}.

\subsection{Constraint Solvers}\label{sec:CS}

Given a logical formula, expressed in a suitable logic, constraint
solvers attempt to find a model that makes the formula true. The
model typically is a match between variables and values. One of the
most famous constraint solvers, based on first-order logic, is Alloy
Analyzer, the verification engine of Alloy~\cite{Alloy}.
In~\cite{sascha2006}, authors propose an approach that exploits
Alloy for modeling and formally verifying fault-tolerant distributed
systems. More precisely they focus on systems that use exception
handling as mechanism for fault tolerance and in particular they
consider systems designed by using Coordinated Atomic Actions
(CAA)~\cite{CAA95}. CAA is a fault-tolerant mechanism that uses
concurrent exception handling and unifies the features of two
complementary concepts: conversation and transaction.
Conversation~\cite{Randell75} is a fault-tolerant technique for
performing coordinated error recovery in a set of participants that
have been designed to interact with each other to provide a specific
service (cooperative concurrency).

\subsection{Testing}\label{sec:testing}

Testing refers to the dynamic verification of a system's behaviour
based on the observation of a selected set of controlled executions,
or test cases~\cite{testingBertolino}. Testing is the main fault
removal technique. 

A real world project involving 34 independent programming teams for
developing program versions of an industry-scale avionics
application is presented in~\cite{testingISSRE03}. Detailed
experimentations are reported to study the nature, source, type,
delectability, and effect of faults uncovered in the programming
versions. A new test generation technique is also presented with an
evaluation of its effectiveness.

Another approach~\cite{vodca06} shows how fault tolerance and
testing can be used to validate component-based systems. Fault
tolerance requirements guide the construction of a fault-tolerant
architecture, which is successively validated with respect to
requirements and submitted to testing.

\subsection{UML-based approaches for modeling and validating dependable
systems}\label{sec:UML}

The approaches considered in this section share the idea of
translating design models into reliability models.

A lot of works and solutions have been proposed in the context of
the European ESPRIT project HIDE~\cite{BondavalliCLMPS01}. This
project aims at the creation of an integrated environment for
designing and verifying dependable systems modeled in UML.
In~\cite{bondavalli99} authors propose automatic transformations
from UML specifications, augmented with additional required
information (i.e., fault occurrence rate, percentage of permanent
faults, etc...), to Petri Net Models.

A modular and hierarchical approach for dependable software
architectures is proposed in~\cite{MPB03}. The language used for
describing software architectures is UML. The approach suggests a
refinement process allowing the description of critical parts of the
model when information becomes available in the following design
phases.

In~\cite{ganesh2002} authors convert UML models to dynamic fault
trees. In this work UML is mainly used as a language for describing
module substitution and error propagation.

\section{Languages and Frameworks}\label{languages}

It is of great importance that engineers could find in their development tools features that help them to deal with the increase of the complexity because of the incorporation of fault-tolerant software techniques into the final software. Each development tool studied in this section helps in separating the code to implement the software system function (as described by its functional specification) from the necessary code to implement
the service restoration (or simply ``recovery''), when a deviation from the correct service was detected (by the implemented error detection technique, of course). The choice of recovery features depends on the classes of faults that to be tolerated. For example, transient faults, which are the faults that eventually disappear without any apparent intervention, can be tolerated by error handling.

Each of the tools that is presented below, allows engineers to solve
this issue and sometimes with several possibilities. The selection
amongst all the proposed solution paths will depend on: costs in
terms of money, processing power (performance), and memory size. But
also on the costs (quantifiable or not) induced by the failure of
the software system, which is the most important one to evaluate
precisely in order to decide on the requirements for fault
tolerance.

This section addresses three types of development environments:
programming languages, fault-tolerant frameworks and advanced
fault-tolerant frameworks. The choice among these three categories
will depend on the complexity of the fault tolerance requirements.

\subsection{Programming Languages Perspectives}\label{PL_perspectives}

Some programming languages incorporate fault tolerance techniques
directly as part of their syntax or indirectly by features that
allow engineers to implement them. One reason for having fault
tolerance support at the programming language level is due to the
increased performance as consequence of the application-specific
knowledge. Another reason is to offer to programmers that need to
comply to standard programming languages the capability to design
and develop more easily fault-tolerant applications.


\subsubsection{Exception Handling}

As stated in the previous section, one of the features to be
provided by a fault tolerance technique is to support separation of
the fault tolerance instructions (for recovery objectives) from the
rest of the software and to activate them automatically, when
necessary. The obvious moment to activate the recovery behaviour is
when it is impossible to finish the operation that the software is
carrying out. An exception is exactly defined as the notification of
the impossibility of finishing an operation and it can be used to
know that the software is going to fail if no action is taken. It
points out that the period of time between the notification of the
exception and the failure of the software can be used to apply the
fault-tolerant instructions in order to keep the software running
and avoid a failure. This is called \emph{Exception handling} (EH).
It is the most popular way used by modern software and it plays a
vital part in the implementation of fault tolerance in software
system.

Nowadays, various exception handling models are part of practical
programming languages like Ada, C++, Eiffel, Java, ML and Smalltalk.
Almost all languages have similar basic types of exceptions and
similar constructs to introduce new exception types and to handle
exceptions (e.g. \emph{try/throw/catch} in Java). Thus, exception handling is a
good technique to implement fault-tolerant sequential programs.

In the context of distributed concurrent software (network of
computing nodes), the situation is different. The exception handling
must be defined according to the semantics of concurrency and
distribution. ERLANG~\cite{Erlang96} is a declarative language for
programming concurrent and distributed software systems with EH
features. This language has primitives which support the creation of
processes (separated unit of computation), the communication between
processes over a network of nodes and the handling of errors when a
failure caused a process to terminate abnormally. The statement
spawn allows a process to create a new process on a remote node.
Whenever a new process is created, the new process will belong to
the same process group as the process that evaluated the spawn
statement. Once a process terminates (normally or abnormally) its
execution, a special signal is sent to all processes which belong to
the same group. The value of one of the parameters that compose the
signal is used to detect if the process terminated abnormally. In
order to avoid propagating an abnormal signal to the other processes
of the group (i.e. to ensure failure containment), the default
behaviour must be changed. This is achieved by using the catch
instruction, which defines a scope to deal with errors occurred on
the monitored expression.

\subsubsection{Atomic Actions}

Fault tolerance in distributed concurrent software systems can also
be achieved using the general concept of atomic actions. A group of
components (participants, threads, processes, objects, etc.) that
cooperate to achieve a joint goal, without information flow between
the group and the rest of the system for the period necessary to
achieve the goal, constitutes an atomic action. These components are
designed to cooperate inside the action, so that they share work and
exchange information in order to complete the action successfully.
Atomicity guarantees that if the action is successfully executed,
then its results and modifications on shared data become visible to
other actions. But if an error is detected, all the components take
part in a cooperative recovery in order to return without changes on
the shared data. These scheme characteristics allows the containment
and recovery to be easily achieved since the error detection,
propagation and recovery all occur within a single atomic action.
Therefore fault tolerance steps can be attached to each of the
atomic action that forms part of the software, independently from
each other. The first fault-tolerant atomic action scheme proposed
was the conversation scheme~\cite{Randell75}. It allows tolerating
design faults by making use of software diversity and participant
rollback. Other schemes including fault tolerance has been proposed
since then and developed as part of programming languages. For
example, Avalon~\cite{Detlefs88}, takes advantage of inheritance to
implement atomic actions in distributed object-oriented
applications. Avalon relies on the Camelot~\cite{Spector88} system
to handle operative-system level details. Much of the Avalon design
has been inspired by Argus~\cite{Liskov88}.

\subsubsection{Reflection and Aspect-Orientation}
Other software technology considered for handling software faults
and which is related to programming languages is
``reflection''~\cite{Randell93}. Reflection is the ability of a
computational system to observe its own execution and, as a result
of that observance, perhaps make changes to that execution. Software
based on reflective facilities is structured into different levels:
the base level and one or more metalevels. Everything in the
implementation and application (the syntax, the semantics, and the
run-time data structures) is ``open" to the programmer for
modification via the metalevels~\cite{Rogers03}. The metalevels can
be used to handle the fault tolerance strategies. Therefore, this
layer structure allows programmers to separate the recovery steps
(part of the metalevels) from necessary steps to reach the
functional goal (part of the base level). The fact that metalevels
can observe the base level computation allows halting its execution
when any deviation is observed (according to some parameter of
reference) to start the recovery.

A generic solution to implement fault tolerance that is or can be
chosen for many  programming languages consists in:

\begin{itemize}
\item extending the programming language with non-standard constructs and semantics;
\item extending the implementation environment underlying the programming language to
provide the functionality, but with an interface expressed using
existing language constructs and semantics;
\item extending the language with specific abstractions, implemented with existing language
constructs and semantics (e.g., abstract data types intended to
support software fault tolerance) perhaps expressed in
well-recognised design patterns;
\item or combinations of the previous approaches.
\end{itemize}


Aspect-orientation has been accepted as a powerful technique for
modularizing crosscutting concerns during software development in
so- called aspects. Similar to reflection, aspect-oriented
techniques provide means to extend a base programs with additional
state, and define additional behaviour that is to be triggered at
well-defined points during the execution of the program. Experience
has shown that aspect-oriented programming is successful in
modularizing even very application-independent, general concerns
such as distribution and concurrency, and examples of using
aspect-orientation to achieve fault tolerance are given
in~\cite{kienzle2002b,soares02,rashid03,kienzleomtt,fabry2005}.

Nevertheless, if complex fault tolerance requirements are given to
your application, choosing programming languages will be risky and
the selection of a framework will then be necessary.

\subsection{Frameworks for Fault Tolerance}\label{FT_frameworks}

According to~\cite{Johnson97} \emph{``a framework is a reusable
design expressed as a set of abstract classes and the way their
instances collaborate. It is a reusable design for all or part of a
software system. By definition, a framework is an object-oriented
design. It doesn't have to be implemented in an object-oriented
language, though it usually is. Large-scale reuse of object-oriented
libraries requires frameworks. The framework provides a context for
the components in the library to be reused.''}

CORBA (Common Object Request Broker Architecture), is a good example
of a framework, which was conceived to provide application
interoperability. Unfortunately, CORBA and other traditional
frameworks cannot often meet the demanding quality of service (QoS)
requirements (including the fault-tolerance ones) for certain
specialised applications. This is why these frameworks are often
extended to include fault tolerance techniques in order to become
predictable and reliable. This is done in FT CORBA
specification~\cite{OCIWEB}. It defines architecture, a set of
services, and associated fault tolerance mechanisms that constitute
a framework for resilient, highly-available, distributed software
systems. Fault tolerance is achieved by features that allow
designers to replicate objects in a transparent way. The set of
several replicas for a specific object defines an object group.
However, a client object is not aware that the server object is
replicated (server object group). Therefore, the client object
invokes methods on the server object group, and the members of the
server object group execute the methods and return their responses
to the client, just like a conventional object.

The same approach has also been followed at the higher levels of
abstraction. For example, the well-known coordination language
Linda~\cite{Gelernter85}, which has been extended to facilitate
programming of fault-tolerant parallel applications.
FT-Linda~\cite{Bakken95}, for instance, is an extension of the
original Linda model, which defines new concepts as stable tuple
spaces (stable TSs), failure tuple and new syntax to achieve atomic
execution of a series of TSs operations. A stable TS represents a
tuple that survives to a failure. This tuple stability is achieved
by replicating the tuple on the multiple hosts that conform the
distributed environment where the software is deployed. FT-Linda
uses monitoring to detect failures. The main type of failure that is
addressed by FT-Linda corresponds to host failure, which means that
the host has been silent for longer than a pre-defined interval.
When such type of failure is detected, the framework automatically
notifies all processes by signalling a failure tuple in a stable TS
available to them. Notice, there must exist a specific process in
the software in charge of watching for a failure tuple and starting
the corresponding recovery process. Atomic execution is denoted by
angle brackets and represents the all-or-none execution of the
group of tuple space operations enclosed by the angle brackets.
There are also some language primitives that allow tuples to be
moved or copied from one TS to another one.

Other extension of Linda are oriented towards mobile applications
using the agent paradigm, is CAMA~\cite{Arief06}. It brings fault
tolerance to mobile agent applications, by the use of a novel
exception handling mechanisms developed for this type of
applications. This exception handling mechanism consists in
attaching application-specified handlers to agents to achieve
recovery. Therefore, to the set of primitives derived from Linda
(e.g. create, delete, put, get, etc.), CAMA adds some primitives to
catch and raise inter-agent exceptions (e.g. raise, check, wait).

Linda has strongly influenced the JavaSpaces~\cite{JavaSpaces}
system design. They are similar in the sense that they store
collections of information (resources) which can be later searchable
by value-based lookup in order to allow members of distributed
software systems to exchange information. JavaSpaces is part of Jini
Network Technology~\cite{Jini}, which is an open architecture that
enables developers to create network-centric service. Jini has a
basic mechanism used for fault-tolerant resource control, which is
referred to as Lease. This mechanism is used to define a holding
interval of time on a resource by the party that requests access on
such resource. The mechanisms notifies (error detection) when the
leases expires, so that actions (recovery) on lease expiration can
be taken for the requestor party.

Even though these framework extensions represent good tools to
implement fault-tolerance they are still ``extensions'' of existing
tools. The next section will present frameworks that have been
designed with the central objective to support the design and
implementation of fault tolerance.

\subsection{Advanced Frameworks for Fault Tolerance}\label{Advanced_FT_frameworks}


Frameworks that have been defined to allow designers/programmers to
develop fault-tolerant software by the implementation of fault
tolerance techniques share, in general, the following
characteristics~\cite{Pullum01}:

\begin{itemize}
\item many details of their implementation are made transparent to the programmers;
\item they provide well-defined interfaces for the definition and implementation of fault tolerance techniques;
\item they are recursive in nature (each component can be seen as a system itself).
\end{itemize}

Arjuna~\cite{Parrington95} programmed in C++ and Java, is one of
those. Arjuna permits the construction of reliable distributed
applications in a relatively transparent manner. Reliability is
achieved through the provision of traditional atomic transaction
mechanisms implemented using only standard language features. It
provides basic capabilities to handle recovery, persistence and
concurrency control to the programmer, while at the same time giving
flexibility to the software by allowing those capabilities to be
refined as required by the demands of the application.

Other similar work is OPTIMA~\cite{Kienzle01}, which is an
object-oriented framework (developed for Ada, Java and AspectJ) that
provides the necessary runtime support for the ``Open Multithreaded
Transactions'' (OMT) model. OMT is a transaction model that provides
features for controlling and structuring not only accesses to
objects, as usually happens in transaction systems, but also threads
taking part in a same transaction in order to perform a joint
activity. The framework provides features to fork and terminate
threads inside a transaction, but restricting their behaviours in
order to guarantee correctness of transaction nesting and isolation
among transactions.

The DRIP framework~\cite{ZorzoStroud99} provides a technological
support to implement DIP~\cite{Zorzo99} models in Java. DIP combines
Multiparty Interactions~\cite{Evangelist89, Joung96} and Exception
Handling~\cite{cristian89} in order to support design dependable
interactions among several processes. As an extension of DRIP, the
CAA-DRIP implementation framework~\cite{Capozucca06} provides a way
to execute Java programs that have been designed using the COALA
design language~\cite{Vachon00} that exploits the Coordinated Atomic
actions (CA actions) concept~\cite{Randell97}.

Other advanced frameworks for fault-tolerance are currently being
designed. For example, the MetaSolve design framework, supporting
dynamic selection and parallel composition of services, has been
proposed for developing dependable systems~\cite{DiMarzo07}. This
approach defines an architectural model, which makes use of
service-oriented architecture features to implement fault tolerance
techniques based on meta-data. Furthermore, the software implemented
using the MetaSolve approach will be able to adapt itself at
run-time in order to provide dynamic fault-tolerance. The ability of
such software to dynamically respond to potentially damaging changes
by adaptation in order to maintain an acceptable level of service is
refereed to as dynamic resilience. The architecture of software that
follows this approach relies on dynamic information about software
components in order to make decisions for dynamic reconfiguration.
Such metadata then will be used in accordance with some resilient
policies ensuring that the desirable dependable requirements are
met.
\section{Contribution of this Book to the Topic }\label{sect2}

This book will contribute to the overall topic of Software
Engineering and Fault Tolerance with nine papers, briefly described
in the following, and categorized according to the three parts
identified at the beginning of this paper:

\begin{itemize}
\item Part A: {\em Fault tolerance engineering: from requirements to code}

    \begin{itemize}
    \item {\em In ``Exploiting Reflection to Enable Scalable and Performant Database
    Replication at the Middleware Level" Jorge Salas, Ricardo
    Jimenez-Peris, Marta Patino-Martinez and Bettina Kemme} introduce a
    design pattern for data base replication using reflection at
    interface level. It permits a clear separation between regular
    function and replication logic. This design pattern allows to obtain
    good performance and scalabilitie properties.

    \item {\em In ``Adding Fault-Tolerance to State Machine-Based Designs", Sandeep
    S. Kulkarni, Anish Arora and Ali Ebnenasir} present a non application
    specific approach for automatic re-engineering of code in order to
    make it fault-tolerant and safety. This generic approach is using
    model based transformations of programs that must be atomic
    concerning accesses to variables.

    \item {\em In ``Replication in Service-Oriented Systems", Johannes Osrael,
    Lorenz Froihofer and Karl M. Goeschka} present a state of the art of
    replication protocols and of replication in service oriented
    architectures supported by middleware's. Then is shows how to
    enhance to existing solutions provided in the service oriented field
    with already known designs for replication in traditional systems.
    \end{itemize}

\item Part B: {\em Verification and validation of fault tolerant
systems}

    \begin{itemize}
    \item {\em In ``Embedded Software Validation Using On-Chip Debugging
    Mechanisms", Juan Pardo, Jos\'{e} Carlos Campelo, Juan Carlos Ruiz and
    Pedro Gil} present how to practically use on-chip debugging to
    perform fault-injection in a non-intrusive way. This portable
    approach offers a verification and validation mean for checking and
    validating the robustness of COTS based embedded systems.

    \item {\em In ``Error Detection in Control Flow of Event-Driven State Based
    Applications", Gergely Pinter and  Istvan Majzik} present a formal
    approach using state-charts to detect two classes of faults: the one
    made during state-chart refinement (using temporal logic
    model-checking) and the one made at implementation (using model
    based testing).

    \item {\em In ``Fault-Tolerant Communication for Distributed Embedded Systems",
    Christian K\"{u}hnel and Maria Spichkova} present a formal specification
    using the FOCUS formal framework of FlexRay and FTCom. It thus
    provides a precise semantics useful for analyzing dependencies and
    for the verification (using Isabelle/HOL) of existing
    implementations.
    \end{itemize}

\item Part C: {\em Languages and Tools for engineering fault
tolerant systems}

    \begin{itemize}
    \item {\em In ``A Model Driven Exception Management Framework", Susan Entwisle
    and Elizabeth Kendall} present a model driven engineering approach to
    the engineering of fault-tolerant systems. An iterative development
    process using the UML 2 modelling language and model transformations
    is proposed. The engineering framework proposes generic
    transformations for exceptions handling strategies thus raising the
    exception handling at higher level of abstraction than only
    implementation.

    \item {\em In ``Runtime Failure Detection and Adaptive Repair for Fault-Tolerant
    Component-Based Applications", Rong Su, Michel Chaudron and Johan
    Lukkien} present formally a fault management mechanism useful for
    systems designed using a component model. Run-time failure are
    detected and the repair strategy is selected using a rule-based
    approach. An adaptive technique is proposed to enhance progressively
    the selection of the repair strategy. A forthcoming development
    framework, Robocop, will provide an implementation of this
    mechanism.

    \item {\em In ``Extending the Applicability of the Neko Framework for the
    Qualitative and Quantitative Validation and Verification of
    Distributed Algorithms", Lorenzo Falai and Andrea Bondavalli} present
    a development framework written in Java allowing rapid prototyping
    of Java distributed algorithms. An import function allows for direct
    integration of C and C++ programs via glue-code. The framework
    offers some techniques for qualitative analysis that can be used
    specifically for the fault tolerance parts of the distributed
    program developed with the framework.
    \end{itemize}

 \end{itemize}

\section*{Acknowledgments}
The book editors wish to thank Andrea Bondavalli and Rogerio de
Lemos for their constructive comments on this introductory chapter
and Alfredo Capozucca  and Joerg Kienzle for their comments and
contribution to Section 6.

\bibliographystyle{ws-procs9x6}
\bibliography{SEFT}
\end{document}